%% file: paper.tex
\documentclass[aps,prl,twocolumn,superscriptaddress]{revtex4-2}

\usepackage{hyperref}
\usepackage{graphicx}
\usepackage{siunitx}
\usepackage{amsmath}
\usepackage{soul}
\usepackage[labelformat=simple]{subcaption}
\usepackage{svg}
\usepackage{enumitem}

\usepackage{dcolumn}
\usepackage[justification=raggedright, singlelinecheck=false]{caption}

\renewcommand{\eqref}[1]{Eq.~(\ref{#1})}

\begin{document}

\include{macros}


\title{Frequency-Time Multiplexing for Near-Deterministic Generation of $n$-Photon Frequency-Bin States}


\author{Alex Fischer}
\affiliation{Sandia National Laboratories, Albuquerque, New Mexico 87123, USA}
\affiliation{Center for Quantum Information and Control, Department of Physics and Astronomy, University of New Mexico,
Albuquerque, New Mexico 87131, USA}

\author{Nathan T. Arnold}
\affiliation{Photon Queue Inc., Champaign, Illinois 61820, USA}

\author{Colin P. Lualdi}
\affiliation{Photon Queue Inc., Champaign, Illinois 61820, USA}

\author{Kelsey Ortiz}
\affiliation{Photon Queue Inc., Champaign, Illinois 61820, USA}

\author{Michael Gehl}
\affiliation{Sandia National Laboratories, Albuquerque, New Mexico 87123, USA}

\author{Paul Davids}
\affiliation{Sandia National Laboratories, Albuquerque, New Mexico 87123, USA}

\author{Kai Shinbrough}
\email{kai.shinbrough@photonqueue.com}
\affiliation{Photon Queue Inc., Champaign, Illinois 61820, USA}

\author{Nils T. Otterstrom}
\email{ntotter@sandia.gov}
\affiliation{Sandia National Laboratories, Albuquerque, New Mexico 87123, USA}


\date{\today}

\begin{abstract}

One of the primary challenges of photonic quantum information processing is the on-demand preparation of multiple single-photon-level quantum states from probabilistic photon pair sources. Motivated by recent developments in frequency-bin-encoded photonic quantum information processing, here we consider active time multiplexing to generate $n$-photon states, where $n$ single photons with $n$ distinct frequencies occupy the same spatiotemporal mode. 
We devise an approach that uses optical quantum memories to manipulate the temporal mode of heralded single photons and an array of fiber Bragg grating reflectors to jointly manipulate the frequency and temporal modes of the photons, overlapping $n$ photons in $n$ separate frequency bins into a single spatiotemporal mode. We calculate multiphoton state generation rates that, accounting for loss, are realistically achievable with commercially available hardware. Using only a single free-space switchable delay loop for an optical quantum memory, this scheme could feasibly produce 8-photon states at an average rate of 1 kHz.

\end{abstract}


\maketitle

\paragraph{Introduction.}

Quantum photonics is a promising platform for applications in quantum computing and information processing due to 
the exquisite coherence and isolation of photonic states from their environment, even at room temperature \cite{linear_optical_quantum_computing_rmp}. However, a major challenge associated with photonic quantum information processing is the high-fidelity, high-rate, and deterministic preparation of indistinguishable single-photon states, which serve as the building blocks for most photonic quantum information processing tasks. This challenge is particularly acute for photonic sources based on room-temperature-compatible probabilistic photon pair sources, which must be driven at low generation probability to limit multipair generation \cite{linear_optical_quantum_computing_rmp}.

The widely accepted solution to this problem relies on \textit{heralding} the generation of a single photon by detection of its conjugate in a photon pair, and \textit{multiplexing} of multiple trial modes into a single output mode in order to approach deterministic generation of single-photon states \cite{jeffrey2004multiplexing,multiplexing_migdall_2002,multiplexing_pittman_2002,kaneda_multiplexing_review_2026}. Many degrees of freedom have been investigated experimentally for multiplexing single photons, including time bin \cite{kaneda_kwiat_2019_temporal_multiplexing,kaneda2015time}, spatial mode \cite{collins_2013_spatial_multiplexing,nunn2013enhancing,glebov2013deterministic}, and frequency bin \cite{frequency_conversion_sinclair_14,frequency_conversion_li_18,joshi2018frequency,frequency_conversion_hiemstra_20}, among others. Here, we consider a new scheme that combines active time-bin multiplexing and passive frequency-bin multiplexing for the preparation of $n$-photon states distributed in frequency. Our preparation of $n$-photon frequency-bin states is motivated by the continued development of frequency-encoded photonic qubits for photonic quantum information processing \cite{lukens_17_frequency_qip,frequency_encoded_photonic_qc_2023,frequency_photons_50ghz,qkd_frequency,on_chip_frequency_photonics}, which provides a scalable route to large-scale photonic quantum information processing.

For dual-rail photonic quantum computation, resource states are produced by interfering multiple single photons that arrive at the same time at separate input ports of an interferometer \cite{bartolucci2021_resource_state_generation}. A standard protocol for Bell-state generation \cite{bartolucci2021_resource_state_generation} requires 4 photons arriving at separate interferometer ports at the same time. More complicated resource state generation will require order tens 
of photons or more \cite{loss_thresholds_psiquantum}. Therefore, a requirement for dual-rail photonic quantum computation is not only generating single-photons in fixed temporal modes, but also generating multiple single photons in identical temporal modes and distinct modes of the degree of freedom that encodes the computational information (e.g., spatial mode, polarization, frequency). These distinct single photons aligned in time will arrive at separate ports of an interferometer, either a traditional spatial interferometer, or more general interferometers that mix other degrees of freedom such as polarization or frequency.

When these computational rails are frequency bins, as we consider in this work, 
the relevant target states consist of multiple photons, each with a distinct, fixed frequency, all occupying the same spatio-temporal mode. For heralded single-photon sources with probability $p$ of generating a single photon in a specific frequency, the probability of generating $n$ photons in $n$ distinct frequencies in the same time bin decays exponentially as $p^n$, where typically $p\sim 0.1$ or less. This probability rapidly vanishes as $n$ grows. Even with $n=6$ 
and a $p=0.1$ photon pair source, the probability of generating enough photons for a GHZ state generator is $\mathcal{O}(10^{-6})$ \cite{bartolucci2021_resource_state_generation}. Porting large $m\times n$ switch networks from the spatial domain \cite{bartolucci2021switch} to the frequency domain is not feasible because low-loss, high-speed, frequency-specific spatial switches or frequency converters are not available. So another strategy must be used to multiplex multiple single photons into different frequency bins at the same time.

\begin{figure*}[ht!]
    \includegraphics[width=\linewidth]{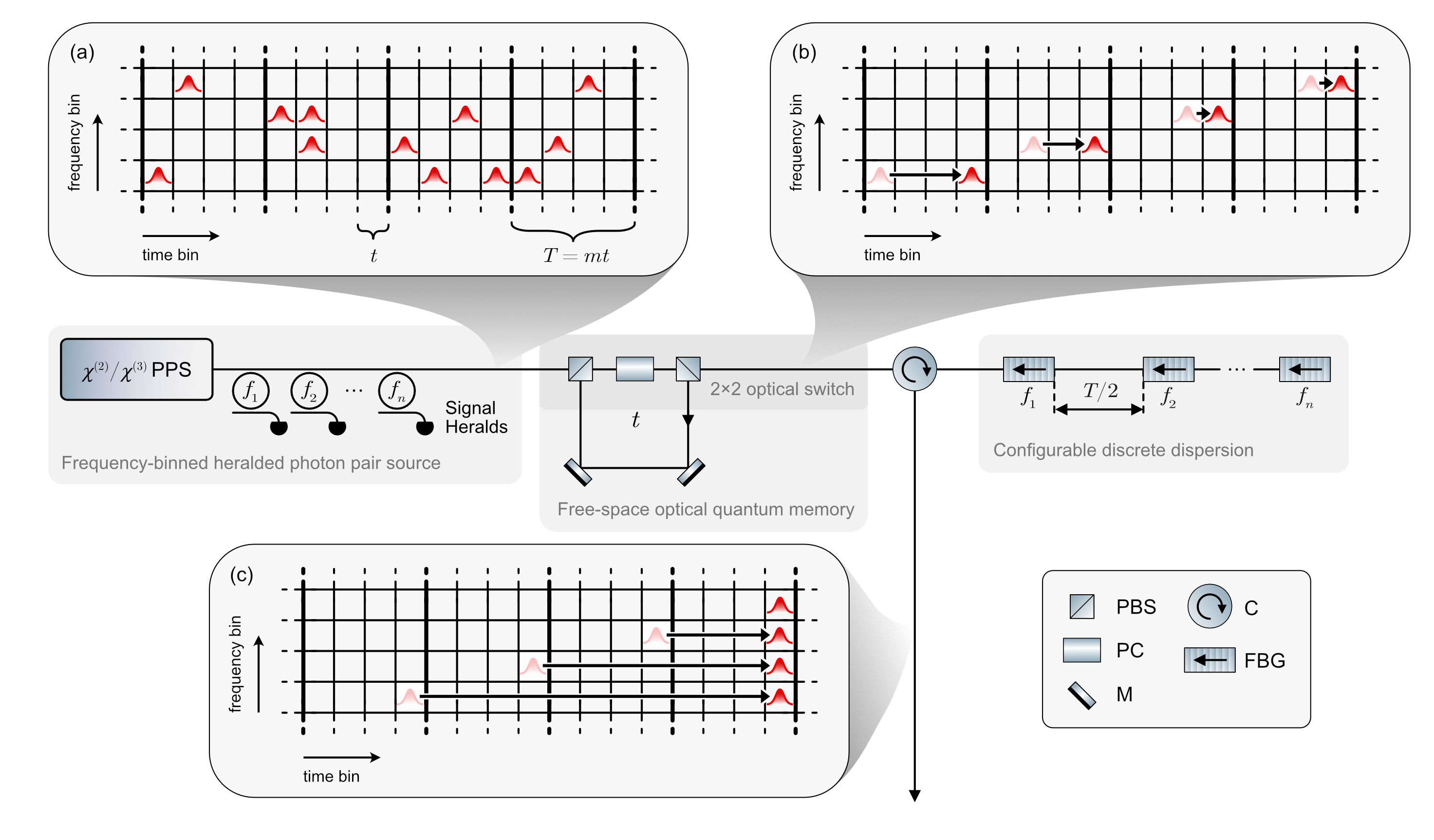}
    \caption{Schematic of our multiplexing scheme.
    A photon pair source (PPS) pumped with a pulsed laser produces signal-idler photon pairs in discrete frequency and time bins. Time bins have length $t$.
    Measuring the signal photons heralds the existence of idler photons in known, random frequency and time bins. An example frequency and time diagram is shown in (a). 
    Bold lines delineate batches of $m$ time bins. 
    The optical quantum memory is shown here as a switchable free-space optical delay. A free-space loop defined by mirrors (M) gets light switched in and out of it with polarizing beamsplitters (PBS) and a Pockels Cell (PC). It delays selected photons such that each batch, corresponding to one frequency bin, has a photon in the last time bin of the batch [see (b)].
    Fiber Bragg grating (FBG) reflectors spaced by $T/2=mt/2$ align photons in each frequency bin into the same time bin [see (c)].
    A circulator (C) redirects the photons reflected from the FBGs to an output port.
    }
    \label{optical_circuit_single_loop}
\end{figure*}

In this work, we address this $p^n$ exponential scaling problem by proposing a multiplexing scheme that uses only broadband, high-speed, free-space optical switches and fixed frequency-dependent delays to produce $n$-photon states with a rate $\mathcal O\left[p/(n\log n)\right]$, not accounting for loss. Here $\mathcal O(\cdots)$ should be understood to mean taking $n\to\infty$ and $p\to 0$. The high-level overview of our multiplexing scheme is that since low-loss, high-speed, and frequency-dependent switches are not feasible, we combine a switchable frequency-independent delay of the kind used in time multiplexing experiments \cite{kaneda_kwiat_2019_temporal_multiplexing} with fixed, non-switchable frequency-dependent delays from spaced fiber Bragg grating reflectors in order to produce switchable frequency-dependent delays. Accounting for loss and assuming optimistic but feasible loss values for commercially available hardware, we show that one could implement this multiplexing scheme to generate 8-photon states at $\sim$ kHz rates, representing a $\sim 2000\times$ improvement relative to no multiplexing. The active elements in this scheme are broadband, and the frequency-dependent elements are passive; we do not use frequency conversion. Our scheme should be understood as a combination of active time multiplexing and passive frequency multiplexing. To the best of our knowledge, this is the first work to propose a method to produce such multiphoton states for frequency-encoded photonic quantum computation, where single photons with different frequencies occupy the same spatio-temporal mode.

\paragraph{Proposed multiplexing scheme.}

An overview of our multiplexing scheme is provided in \figref{optical_circuit_single_loop}. A photon pair source pumped with a pulsed laser produces correlated signal-idler photon pairs in discrete frequency and time bins. The discrete-frequency-bins nature of the photon-pair source is necessary for frequency-bin photonic quantum computation \cite{lukens_17_frequency_qip}. Sources based on spontaneous four-wave mixing can generate photon pairs in these discrete frequency bins automatically \cite{sfwm_photon_pair_source_liscidini_19}, while sources based on spontaneous parametric down conversion require filtering or other modifications to produce photon pairs in discrete frequency bins \cite{spdc_pair_source_frequency_olislager_10, spdc_source_frequency_morrison_22}.

Photon-number-resolving and frequency-bin-resolving measurements of the signal photons herald the presence of single idler photons randomly distributed in frequency and time bins [see Fig.~\ref{optical_circuit_single_loop}(a)]. We assume photon-number-resolving detectors are available because they are necessary for linear optical quantum computation, which is the main expected application of this work. Therefore we do not consider multiphoton errors.

We split up the stream of generated time bins into \textit{batches}, which are groups of $m$ time bins, and we associate each batch with one frequency bin. For that batch, we apply the time-multiplexing scheme of Ref.~\cite{kaneda_kwiat_2019_temporal_multiplexing} and delay the last photon in the batch with the correct frequency to the last time bin in the batch [see Fig.~\ref{optical_circuit_single_loop}(b)], neglecting the rest of the photons in the batch in different frequency and time bins. 
We utilize the frequency degree of freedom to transform the temporal configuration of the photons from the fixed configuration in Fig.~\ref{optical_circuit_single_loop}(b), to one in which all $n$ photons overlap in time. A frequency-dependent delay consisting of an array of fiber Bragg grating (FBG) reflectors transforms the temporal structure of the state such that all photons are aligned in a single time bin [see Fig.~\ref{optical_circuit_single_loop}(c)]. The array of FBG reflectors acts as a discrete dispersive element, causing light in different frequency bins to travel a longer distance proportional to its frequency difference from a reference frequency bin. The FBGs in the array are spaced by time $mt/2$ apart, where $m$ is the number of time bins per batch, and $t$ is the length between time bins. This $mt/2$ spacing is half the length of a batch, so that sequential frequency photons have a round-trip path length difference equal to the length of one batch. A circulator redirects the reflected photons into an output port.

\paragraph{$n$-Photon Rates.}

\begin{figure}
    \includegraphics[width=\linewidth]{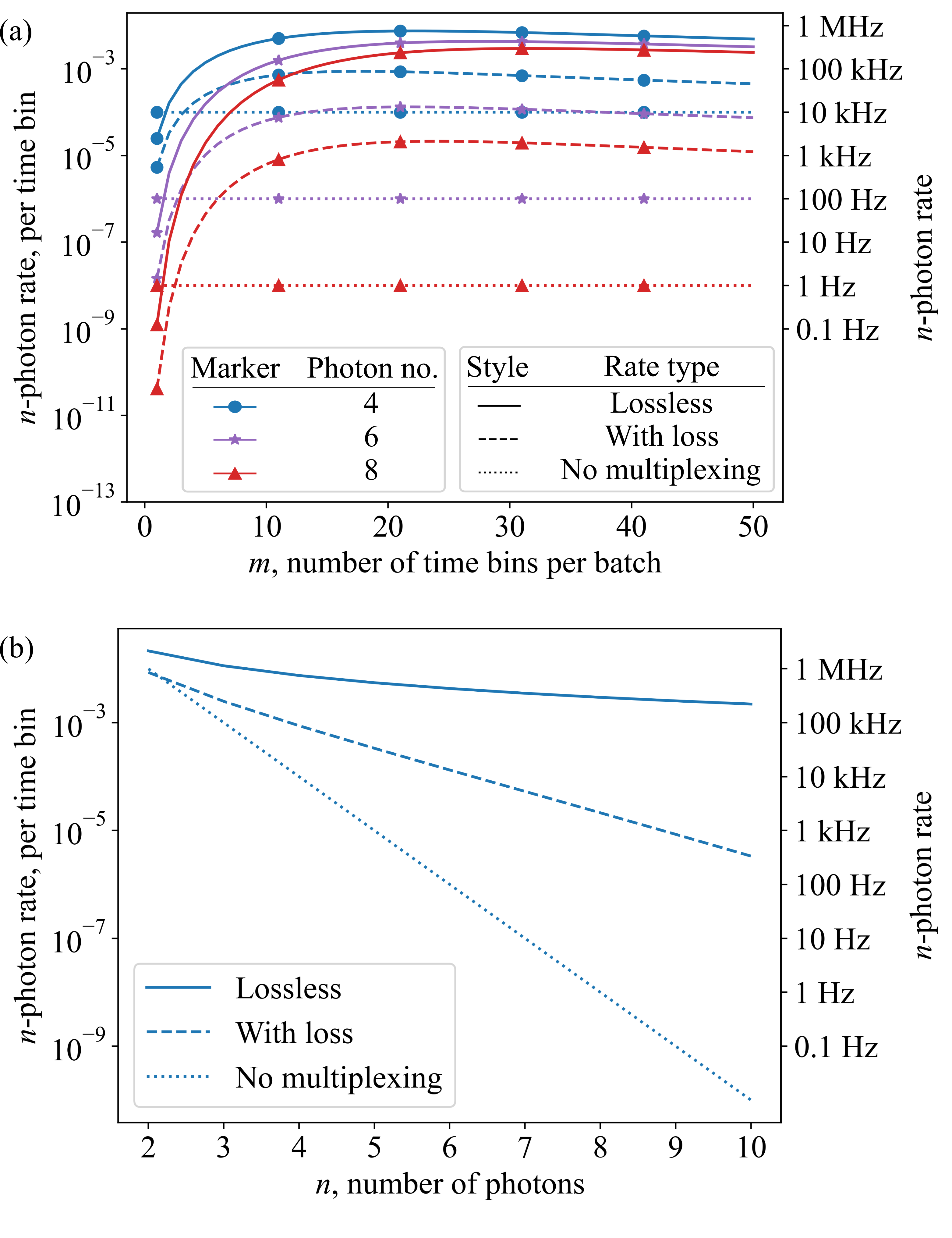}
    \caption{$n$-photon generation rates.
    (a) $n$-photon generation rates, with (dashed) and without (solid) loss, with and without (dotted) multiplexing. We show both dimensionless rates that are the expected number of multiphoton events per time bin (left axis), and dimensionful rates that assume a 10 ns time bin length (right axis).
    (b) Maximum $n$-photon generation rates, with and without accounting for loss, with and without multiplexing, as a function of number of photons to generate.}\label{rates_single_loop}
\end{figure}

\begin{figure}[ht!]
    \includegraphics[width=\linewidth]{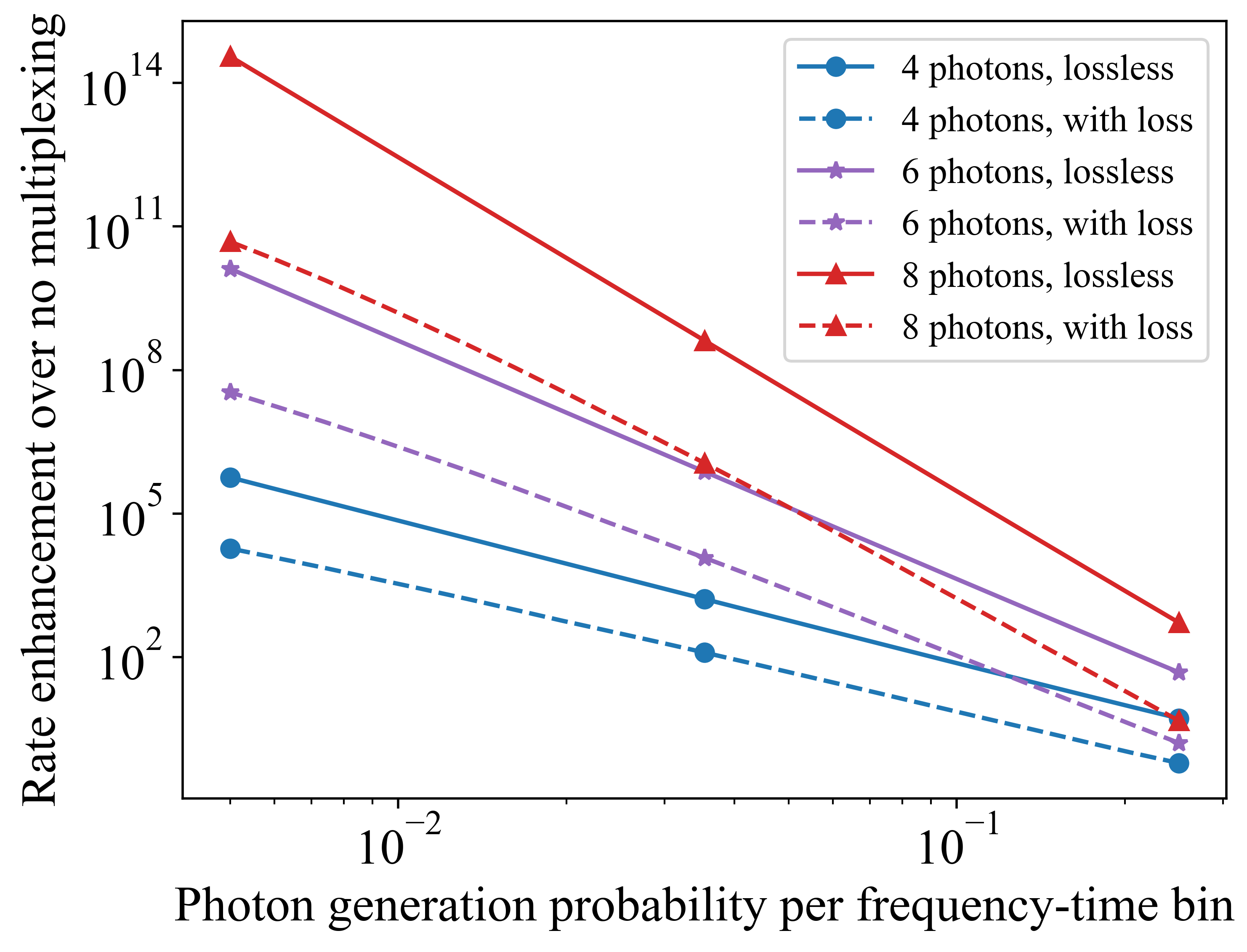}
    \caption{Ratio of $n$-photon generation rates with our multiplexing scheme versus without multiplexing. Here we assume 100-, 10-, and 1-timestep length storage loops, to enable storing light for hundreds of timesteps without high loss. For each photon number $n\in\{4,6,8\}$, we show the ratio of the rate at the choice of number of time bins per batch $m$ with maximal rate, to the rate without multiplexing $p^n$, as a function of the probability of generating a photon per frequency-time bin $p$.}\label{rate_enhancement}
\end{figure}

The temporal duration of our multiplexing scheme is $mn$ time bins, where $m$ is the number of time bins per batch, and $n$ is the number of frequency bins (and therefore batches) and the number of photons to be outputted in the same time bin. Not accounting for loss, the success probability of this scheme is $\left[1-(1-p)^m\right]^n$, where $p$ is the probability of the frequency-binned heralded single photon source producing a photon in any particular frequency and time bin. The entire multiplexing scheme takes $mn$ time bins, where $m$ is the number of time bins per batch, and $n$ is the number of photons (equivalently, number of batches). Thus the rate of $n$-photon state generation, not accounting for loss (equivalently, the rate of $n$-photon \emph{heralding}) is that success probability $\left[1-(1-p)^m\right]^n$ over the length of time of the entire scheme $mn$, which is 
\begin{align}
    R=\frac{\left[1-(1-p)^m\right]^n}{mn}\label{rate_lossless}
\end{align}
in units of reciprocal time bin. We plot these lossless $n$-photon generation rates for $n=4, 6, 8$ in \figref{rates_single_loop}(a). We assume a $p=0.1$ photon generation probability per time and frequency bin, here and everywhere else unless stated otherwise. The left axis shows dimensionless rates per time bin, and the right axis shows these rates assuming a $t=10$ ns time bin length.

Setting $m$, the number of time bins per batch, to $\log_{1/(1-p)}(n/\epsilon)$, ensures that the success probability $\left[1-(1-p)^m\right]^n$ is $\geq 1-\epsilon$. Thus, for a choice of failure probability $\epsilon$, an achievable rate with this scheme is
\begin{align}
    R = \frac{1-\epsilon}{n\log_{\frac{1}{1-p}}\frac{n}{\epsilon}}
\end{align}
in units of reciprocal time bins. As $n\to\infty$ and $p\to 0$, this rate scales as $\mathcal O\left[p/(n\log n)\right]$,
which is much more favorable than the $p^n$ rate without multiplexing.

For $m=1$, the $n$-photon lossless rates are $p^n/n$. This is a factor $n$ smaller than the rate without multiplexing ($p^n$), as the time-bin realignment procedure [shown in \figref{optical_circuit_single_loop}(c)] increases the protocol duration without increasing the success rate.

The lossless rates in \figref{rates_single_loop}(a) shows that there exists an optimal $m$ for each target photon number $n$. Increasing $m$ increases the probability that there will be at least one photon per batch; however, once this probability saturates near 1, increasing $m$ increases the total time the multiplexing scheme takes (time $nmt$), which decreases the multiphoton state generation rate. Therefore, each photon number $n$ will have some optimal number of time bins per batch $m$ that achieves an optimal rate. We plot these optimal rates as a function of photon number in \figref{rates_single_loop}(b).

$n$-photon rates accounting for loss require a slightly more complicated calculation than the simple expression \eqref{rate_lossless}, which we do using optimistic but feasible loss values for commercially available hardware, listed in Table \ref{loss_table}. With a photon-generation probability $p$ per frequency-time bin, and $m$ time bins per batch, the probability of generating an $n$ photon state in $n$ frequency bins without losing any photons is
\begin{align}
    \prod_{\nu=0}^{n-1}\sum_{\tau=0}^{m-1} p(1-p)^{\tau}P(m;\nu,\tau)\label{lossy_success_probability}
\end{align}
where $P(m;\nu,\tau)$ is the probability that a photon in frequency bin $\nu$ and time bin $\tau$ within its batch is not lost as it traverses the entire setup of \figref{optical_circuit_single_loop} with $m$ time bins per batch. $P(m;\nu,\tau)$ can be straightforwardly calculated by adding multiples of the loss values in Table \ref{loss_table}. The $n$-photon generation rate is the $n$-photon success probability accounting for loss [\eqref{lossy_success_probability}] divided by the time the entire multiplexing scheme takes ($mn$ time bins). This rate is
\begin{align}
    \frac{1}{mn}\prod_{\nu=0}^{n-1}\sum_{\tau=0}^{m-1} p(1-p)^{\tau}P(m;\nu,\tau)\label{lossy_success_rate}
\end{align}
in units of inverse time bin. We plot these $n$-photon rates for $n=4, 6, 8$ accounting for loss in \figref{rates_single_loop}(b). As with the lossless rates (solid lines), the lossy rates (dashed lines) show that for each photon number $n$, there is an optimal number of time bins per batch $m$ that maximizes the $n$-photon generation rate. Decreasing $m$ below this optimal value decreases the probability that sufficient photons exist to occupy all frequency bins, reducing the rate. Increasing $m$ above this optimal value increases the time the entire multiplexing scheme takes, which also decreases the rate. \figref{rates_single_loop}(a) shows that the optimal $m$ is generally different between the lossless and lossy cases. \figref{rates_single_loop}(b) plots this optimal rate achieved by optimizing over $m$ in the lossless and lossy cases, as well as the $p^n$ rate without multiplexing.

We do not consider multiphoton errors, because we assume that photon-number-resolving detectors are used, as they are necessary for linear optical quantum computing. We also do not consider loss associated with extracting the frequency-binned signals from the source, as identical losses would be incurred without multiplexing, and we directly compare the rates achievable with our scheme to those achievable without multiplexing. In the lossy case, one should view the photon generation probability $p$ as already pricing in the losses associated with the frequency-binned source, which are the same with or without multiplexing.


For photon generation probabilities $\sim 0.01$ or less, this multiplexing scheme is not feasible, as $\sim$ 1\% loss per optical switch means photons will likely be lost after hundreds of passes around the one-timestep storage loop. In order to use this scheme with low photon generation probabilities, one could use multiple storage loops with successively longer storage times, to store photons for many timesteps with fewer passes through lossy switches. In \figref{rate_enhancement}, we show rates achievable with smaller photon generation probabilities $p$, with 100-, 10-, and 1-timestep storage loops as in \cite{photonqueue_memory}. We show these rates as a ratio of rates with multiplexing to rates without multiplexing $p^n$, to emphasize the many orders of magnitude improvement in multiphoton rates with small photon generation probabilities.

In the end matter, we detail the loss assumptions that we use when calculating $n$-photon rates accounting for loss. We also consider $n$-photon generation where the $n$ photons occupy any $n$ distinct frequency bins out of $2n$ possible bins. This is relevant for $2n$ mode/$n$ photon setups for $n/2$-photon resource state generation \cite{bartolucci2021_resource_state_generation}. This setup allows significantly higher $n$-photon state generation rates, at the cost of requiring dynamically switching what frequency-domain beamsplitters are applied to the resulting states. Calculating rates in the $n$ photons/$2n$ frequency bins case requires stochastic simulations of photon generation times and loss, as there is not a simple expression for the probability of generating $n$ photons to evaluate.

\paragraph{Conclusion.}

We present a multiplexing scheme for generating quantum optical states with multiple single photons occupying distinct frequency bins. These states are a necessary ingredient for frequency-bin photonic quantum computation. This scheme is implementable on commercially available hardware, with losses low enough that appreciable rates of 10-photon states are possible. Time is the only actively multiplexed degree of freedom, and we require no nonlinear frequency conversion as is frequently the case in frequency multiplexing schemes \cite{frequency_conversion_sinclair_14,frequency_conversion_li_18,joshi2018frequency,frequency_conversion_hiemstra_20}.

\paragraph{Acknowledgments.}
\begin{acknowledgments}
This material is based upon work supported by the Laboratory Directed Research and Development program at Sandia National Laboratories. Sandia National Laboratories is a multi-program laboratory managed and operated by National Technology and Engineering Solutions of Sandia, LLC.,~a wholly owned subsidiary of Honeywell International, Inc., for the U.S. Department of Energy's National Nuclear Security Administration under contract DE-NA0003525.
\end{acknowledgments}

\bibliography{bib}

\clearpage
\newpage

\section{End matter}

\paragraph{Loss assumptions.\label{loss_section}}

We calculate multiphoton generation rates accounting for loss using optimistic but feasible loss values for commercially available hardware.  These loss values are:
\begin{itemize}[noitemsep]
    \item Loss per memory loops. This is the loss associated with one pass through a 1-, 10-, or 100-timestep memory loop.
    \item Fiber Bragg grating (FBG) transmission loss. This is the loss associated with passing through a FBG.
    \item FBG reflection loss. This is the loss associated with light being reflected by a FBG.
    \item Loss per fiber delay timestep. This is the loss associated with light traveling one timestep (10 ns) in polarization-maintaining optical fiber.
    \item Circulator loss. Light passes through the circulator twice.
    \item Miscellaneous losses. This is primarily fiber coupling loss ($\sim 7\%$ achievable with GRIN lens collimators), and other losses associated with one mandatory pass through all switches and other optics.
\end{itemize}
The loss values we use in modeling the multiphoton state generation rates are in Table~\ref{loss_table}. We do not consider multiphoton errors, as we assume efficient number-resolving detectors (necessary for linear-optical quantum computation) are available. And we do not consider loss associated with the frequency-binned photon pair source, as identical losses would be incurred without multiplexing, and the ratio of rates with multiplexing and without multiplexing is the relevant figure of merit here, and because their construction is addressed elsewhere \cite{spdc_pair_source_frequency_olislager_10,spdc_source_frequency_morrison_22,sfwm_photon_pair_source_liscidini_19} and out of scope for this work.

\begin{table}
    \caption{Loss values for components in the multiplexing scheme that we use when calculating multiphoton state generation rates accounting for loss. These values are the best estimates for loss of commercially available parts that can be used to construct the setup in \figref{optical_circuit_single_loop}.}
    \label{loss_table}
    \begin{ruledtabular}
    \begin{tabular}{lldd}
        Component & \multicolumn{1}{c}{Ref.} & \multicolumn{1}{c}{Loss (dB)} & \multicolumn{1}{c}{Loss (\%)} \\\hline
        Loss per 1 timestep \\ (10 ns) memory loop & \cite{five_nines_loop_loss,switch_loss} & 0.106 & 2.41 \\\hline
        Loss per 10 timestep \\ (100 ns) memory loop & \cite{five_nines_loop_loss,switch_loss} & 0.110 & 2.50 \\\hline
        Loss per 100 timestep \\ (1000 ns) memory loop & \cite{five_nines_loop_loss,switch_loss} & 0.149 & 3.37 \\\hline
        FBG transmission loss & \cite{indie_private_communication} & 0.0436 & 1.00 \\\hline
        FBG reflection loss & \cite{indie_private_communication} & 0.0436 & 1.00 \\\hline
        Loss per 10 ns timestep \\ in fiber & \cite{pm_fiber_loss} & 0.00102 & 0.0235 \\\hline
        Circulator loss & \cite{circulator_loss} & 0.500 & 10.9 \\\hline
        Misc.~losses, \\ one memory loop & \cite{switch_loss,photonqueue_private_communication} & 0.510 & 11.1 \\\hline
        Misc.~losses, \\ two memory loops & \cite{switch_loss,photonqueue_private_communication} & 0.720 & 15.3 \\\hline
        Misc.~losses, \\ three memory loops & \cite{switch_loss,photonqueue_private_communication} & 0.931 & 19.3 \\
    \end{tabular}
    \end{ruledtabular}
\end{table}




\paragraph{Partially filled frequency bins.}

\begin{figure}
    \includesvg[width=\linewidth]{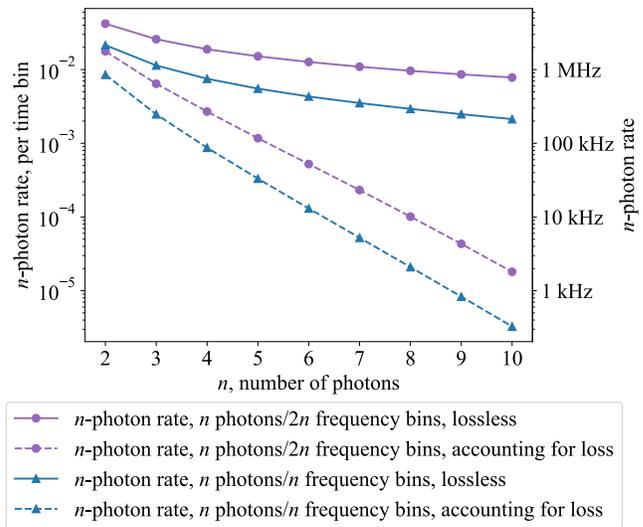}

    \caption{Multiphoton state generation rate for the $n$-photon, $2n$ frequency bins setting, and for the standard $n$-photon, $n$ frequency bins setting. 
    }\label{rate_function_of_num_photons_2n_freq_bins}
\end{figure}

The standard linear optical Bell state generator uses 4 single photons occupying distinct modes and 4 vacuum ancilla modes. And more generally, $n$-GHZ states can be created with $2n$ single photons occupying distinct modes and $2n$ vacuum modes \cite{bartolucci2021_resource_state_generation}. Therefore, useful inputs for linear optical resource state generation do not have to be states with $n$ photons occupying $n$ fixed modes, but rather, $n$ photons occupying any $n$ out of a fixed set of $2n$ modes. The $n$ unoccupied modes will then be the ancilla modes used in resource state generation.

Note that accepting any state where $n$ photons occupy any $n$ modes out of some set of $2n$ modes requires dynamically switching photons between modes, or equivalently, dynamically switching which modes get mixed with beamsplitters. For spatial mode linear optical quantum computing, this may be infeasible, as fast switches are lossy, so switch depth is to be kept at an absolute minimum. For frequency-bin linear optical quantum computing, dynamically altering which frequency bins are mixed with frequency-domain beamsplitters can be done by altering the action of the phase modulators and pulse shapers that make up the frequency-domain beamsplitters. The only extra components that would be added to the photons' paths to dynamically alter the linear optical transformation applied to them would be extra length of optical fiber to have time to send the appropriate signals to the phase modulators and pulse shapers.

Therefore, we also calculate the rates of generating $n$-photon states, where the $n$ photons occupy any $n$ distinct frequency bins out of $2n$ total bins. We see a modest improvement in multiphoton state generation rates in this setting; see \figref{rate_function_of_num_photons_2n_freq_bins}.

For the $n$ photons/$2n$ frequency bins rates, there are not simple to evaluate expressions for multiphoton rates like \eqref{rate_lossless} and \eqref{lossy_success_rate}. Instead, we calculate rates with and without accounting for loss via Monte Carlo sampling. Specifically, we calculate $n$ photons/$2n$ frequency bins rates as follows:
\begin{enumerate}
    \item Take 1,000,000 samples of random configurations of photons appearing with probability $p$ in a $2n$ frequency bin by $2nm$ time bin grid [as in \figref{optical_circuit_single_loop}(a)].
    \item For each configuration of photons in frequency and time, determine if there is a memory switching schedule to fill with photons $n$ of the $2n$ time bins at the end of each batch [filled in \figref{optical_circuit_single_loop}(b)].\label{step_lossless_probability}
    \item Calculate the probability that no photons are lost according to the loss values in Table \ref{loss_table}.\label{step_lossy_probability}
    \item Calculate the success probabilities with and without accounting for loss by averaging the probabilities in steps \ref{step_lossless_probability} and \ref{step_lossy_probability}. Convert success probabilities to rates by dividing by $2nm$.
\end{enumerate}

To find the maximum achievable rate for any given photon number $n$, we maximize over all numbers of time bins per batch $m$ as in \figref{rates_single_loop}(b).

\end{document}

%% file: macros.tex
\newcommand{\figref}[1]{Fig.~\ref{#1}}

\newcommand{\bra}[1]{\left\langle #1\right|}
\newcommand{\ket}[1]{\left| #1\right\rangle}
\newcommand{\braket}[2]{\left\langle #1 \,\right| \!\left. #2\right\rangle}
\newcommand{\brakett}[2]{\left\langle #1 \!\right. \left|\, #2\right\rangle}
\newcommand{\brakettt}[3]{\left\langle #1 \right| #2 \left| #3 \right\rangle}
\newcommand{\ketbra}[1]{\ket{#1}\bra{#1}}

\newcommand{\colvec}[3]{\begin{bmatrix}#1 \\ #2 \\ #3\end{bmatrix}}
\newcommand{\colvecc}[2]{\begin{bmatrix}#1 \\ #2 \end{bmatrix}}
\newcommand{\rowvec}[3]{\begin{bmatrix}#1 & #2 & #3\end{bmatrix}}
\newcommand{\rowvecc}[2]{\begin{bmatrix}#1 & #2\end{bmatrix}}
\newcommand{\mat}[4]{\begin{bmatrix}#1&#2\\#3&#4\end{bmatrix}}
\newcommand{\matd}[4]{\begin{vmatrix}#1&#2\\#3&#4\end{vmatrix}}
\newcommand{\diagg}[2]{\mat{#1}{0}{0}{#2}}
\newcommand{\opnorm}[1]{\left|\left|#1\right|\right|_{\infty}}

\newcommand{\re}[1]{\operatorname{Re}\left(#1\right)}
\newcommand{\im}[1]{\operatorname{Im}\left(#1\right)}
\newcommand{\tr}[1]{\operatorname{Tr}\left[#1\right]}

\newcommand{\sigx}{\mat{0}{1}{1}{0}}
\newcommand{\sigy}{\mat{0}{-i}{i}{0}}
\newcommand{\sigz}{\mat{1}{0}{0}{-1}}
\newcommand{\hmat}{\frac{1}{\sqrt{2}}\mat{1}{1}{1}{-1}}

\newcommand{\expect}[1]{\left\langle#1\right\rangle}
\newcommand{\var}[1]{\operatorname{var}\left(#1\right)}
\newcommand{\infint}{\int_{-\infty}^{\infty}}
\newcommand{\gen}[1]{\left\langle#1\right\rangle}

\newcommand{\expp}[1]{\exp\left(#1\right)}

\newcommand{\even}{\operatorname{even}}
\newcommand{\odd}{\operatorname{odd}}
\newcommand{\purity}{\operatorname{purity}}
\newcommand{\vect}[1]{\operatorname{Vec}\left(#1\right)}
\newcommand{\vc}[1]{\boldsymbol{\mathbf{#1}}}
\newcommand{\ketbar}[1]{\ket{\overline{#1}}}
\newcommand{\cnot}{\mathrm{CNOT}}
\newcommand{\lcnot}{\overline{\mathrm{CNOT}}}
\newcommand{\lx}{\bar{X}}
\newcommand{\lz}{\bar{Z}}
\newcommand{\lh}{\bar{H}}
\newcommand{\ls}{\bar{S}}

\newcommand{\pr}[1]{\operatorname{Pr}\left(#1\right)}
\newcommand{\sign}[1]{\mathrm{sign}\left(#1\right)}

\newcommand{\ie}{i.e., }
\newcommand{\eg}{e.g., }

\newcommand{\del}{\vc\nabla}

\newcommand{\todo}{{\Large\textcolor{red}{\textbf{TODO}}}}